\documentclass[twocolumn,amsmath,amssymb,prb]{revtex4-1}
\usepackage{graphicx}% Include figure files
\usepackage{comment}

\begin{document}
\title{Quantum Oscillations in Cu$_x$Bi$_2$Se$_3$ in High Magnetic Fields
}
\author{B. J. Lawson$^1$, G. Li$^1$, F.  Yu$^1$, T. Asaba$^1$, C. Tinsman$^1$, T. Gao$^1$, W. Wang$^1$, Y. S. Hor$^2$, and Lu Li$^1$}
\affiliation{
$^1$Department of Physics, University of Michigan, Ann Arbor, MI  48109, USA\\
$^2$Department of Physics, Missouri University of Science and Technology, Rolla, Missouri 65409, USA}

\date{\today}
\begin{abstract}
Cu$_x$Bi$_2$Se$_3$ has drawn much attention as the leading candidate to be the first topological superconductor and the realization of coveted Majorana particles in a condensed matter system. However, there has been increasing controversy about the nature of its superconducting phase. This study sheds light on present ambiguity in the normal state electronic state,  by providing a complete look at the quantum oscillations in magnetization in Cu$_x$Bi$_2$Se$_3$ at intense high fields up to 31T.  Our study focuses on the angular dependence of the quantum oscillation pattern in a low carrier concentration. As magnetic field tilts from along the crystalline c-axis to ab-plane, the change of the oscillation period follows the prediction of the ellipsoidal Fermi surface. As the doping level changes, the 3D Fermi surface is found to transform into quasi-cylindrical at high carrier density.  Such a transition is potentially a Lifshitz transition of the electronic state in Cu$_x$Bi$_2$Se$_3$. 
\end{abstract}

%\pacs{74.25.Dw, 74.25.Ha, 74.72.Hs}

\maketitle                   % Produces the title

\section{Introduction}
Topological superconductor is a novel phase of matter that has been theoretically predicted but yet to be experimentally verified. Among topological materials, topological superconductivity is especially interesting because it is a platform to realize Majorana particles - an elusive particle that is its own antiparticle. Furthermore, topological superconductors have been proposed as a platform for topological quantum computation.~\cite{FuKane, WilczekNat, MooreNature} The robustness of the topological surface states makes this avenue an attractive alternative to traditional methods for realizing quantum computation.~\cite{Kitaev}

A topological superconductor must have a full superconducting gap in the bulk with odd parity pairing, and the Fermi surface must enclose an odd number of time reversal invariant momenta in the Brillouin zone, i.e. the Fermi surface must contain an odd number of high symmetry points such as $\Gamma$, Z, X, etc. It also has a topologically protected gapless surface state with Majorana fermions.~\cite{FuKane} Cu$_x$Bi$_2$Se$_3$ has been proposed as a leading candidate for topological superconductivity~\cite{FuBerg} and has sparked a lot of interest. Experiments have shown that by intercalating Cu between Se layers in known topological insulator Bi$_2$Se$_3$ the compound becomes superconducting at 3.8 K.~\cite{Hor104}

Cu$_x$Bi$_2$Se$_3$ has been confirmed to be a bulk superconductor with a full pairing gap by specific heat measurement.~\cite{Kriener} There are some reports of surface Andreev bound states through the observation of Zero Bias Conductance Peak (ZBCP)~\cite{Sasaki}, but other reports that the ZBCP can be removed with gating.~\cite{HPeng}  Recent works using scanning tunneling spectroscopy also did not observed the ZBCP.~\cite{Levy}  ARPES measurements have argued against the topological superconducting mechanism in Cu$_x$Bi$_2$Se$_3$ by reporting an even number of time reversal invariant momenta in the Brillouin zone.~\cite{Lahoud} Both ARPES and quantum oscillation experiments show a Dirac dispersion in Cu$_x$Bi$_2$Se$_3$ - a characteristic feature of topological systems.~\cite{Wray, Lawson} The continual interest in Cu$_x$Bi$_2$Se$_3$ and surmounting controversy of its exotic phase motivates this study for a more complete look at quantum oscillations in magnetization. This work is a continuation and expansion on our previous study of the de Haas-van Alphen effect in Cu$_{0.25}$Bi$_2$Se$_3$~\cite{Lawson} and now includes several samples at a variety of doping levels and complete angular dependence.

From mapping out the Fermi surface, we reveal a closed ellipsoidal Fermi surface that becomes increasingly elongated with increased carrier density. At high carrier concentration, the Fermi Surface crosses the Brillouin Zone boundary and becomes open and quasi-cylindrical. Amplitude damping analysis reveals a strongly anisotropic effective mass. The slope of the energy-momentum dispersion is unchanged with increased Fermi momentum confirming a linear, Dirac-like band structure in Cu$_x$Bi$_2$Se$_3$. The manuscript will first introduce torque magnetometry which we use to resolve quantum oscillations, second it will discuss the results of the angular dependence of the quantum oscillation frequencies, then it will cover the various damping mechanisms of the quantum oscillation amplitude and the parameters extracted from that analysis.

\section{Experiment}
Single crystals of Cu$_x$Bi$_2$Se$_3$ were grown by melting stoichiometric mixtures of high purity elements Bi (99.999\%), Cu (99.99\%), and Se (99.999\%) in a sealed evacuated quartz tube then slowly cooling the mixture from 850$^{\circ}$C down to 620$^{\circ}$C at which point the crystal was quenched in cold water. The doping level was determined according to the mole ratio of the reactants used in the crystal growth, but the nominal doping did not end up corresponding with the measured carrier concentration leaving the precise number under suspicion. Therefore, in this study, we look at how parameters evolve with increased carrier concentration rather than the unreliable nominal doping. The samples used in the study were cut out of large boule of crystals. They are generally black, and the typical size is about 5 mm $\times$ 2 mm $\times$ 0.5 mm.

Quantum oscillations are used to resolve Fermi Surface geometry and to discover electronic properties of topological materials. Oscillations in magnetization, the de Haas-van Alphen effect (dHvA effect), arise from the quantization of the Fermi Surface into Landau Levels.

To measure quantum oscillations in magnetization, $M$, we employed a highly sensitive torque magnetometry method. Torque mangetometry measures the magnetic susceptibility anisotropy of the sample by putting the sample in a tilted magnetic field, $H$, where both $H$ and $M$ are confined to the x-z plane. The torque is then given by $\mathbf{\tau}=\mathbf{M}\times\mathbf{H}=(M_{z}H_{x}-M_{x}H_{z})\mathbf{j}$ and $|\tau|=\chi_{z}H_{z}H_{x}-\chi_{x}H_{x}H_{z}=\Delta\chi H^{2}\sin\phi \cos\phi$, where $\phi$ is the tilt angle of the magnetic field $\mathbf{H}$ away from the crystalline $\hat{c}$ axis and $\Delta\chi=\chi_{z}-\chi_{x}$.

We glue the sample to the head of a thin film cantilever. Both brass cantilevers and Kapton cantilevers with a metalized surface were used. The thinner 0.001 inch brass cantilevers with a higher Young’s modulus and the thicker 0.003 inch Katpon thin films with a lower Young’s modulus offer different spring constants that can provide a balance between strength for heavier samples and sensitivity. The magnetic torque was tracked by measuring the capacitance between the metal surface of the cantilever and a thin gold film underneath. An example of oscillations in the torque data after background subtraction is shown in Fig. \ref{figTorque} with a schematic of the experimental setup in the upper right corner. Oscillations arise from Landau Level quantization. The frequency of this oscillation is proportional to the cross section of the Fermi Surface, $A$, by the Onsager relation:
\begin{equation}
F_{s} = \frac{\hbar}{2\pi e} A.
\end{equation}

To further analyze the oscillation torque pattern, a polynomial background is subtracted from the $\tau - H$ curve to get the oscillatory torque $\tau_{osc}$. A Fast Fourier Transform (FFT) of the oscillatory $\tau_{osc}$ vs. $1/\mu_0H$ is given in the lower left inset of Fig. \ref{figTorque} revealing a single Fermi pocket.

\section{Results}
\begin{figure}[t]
\includegraphics[width= 3.5 in ]{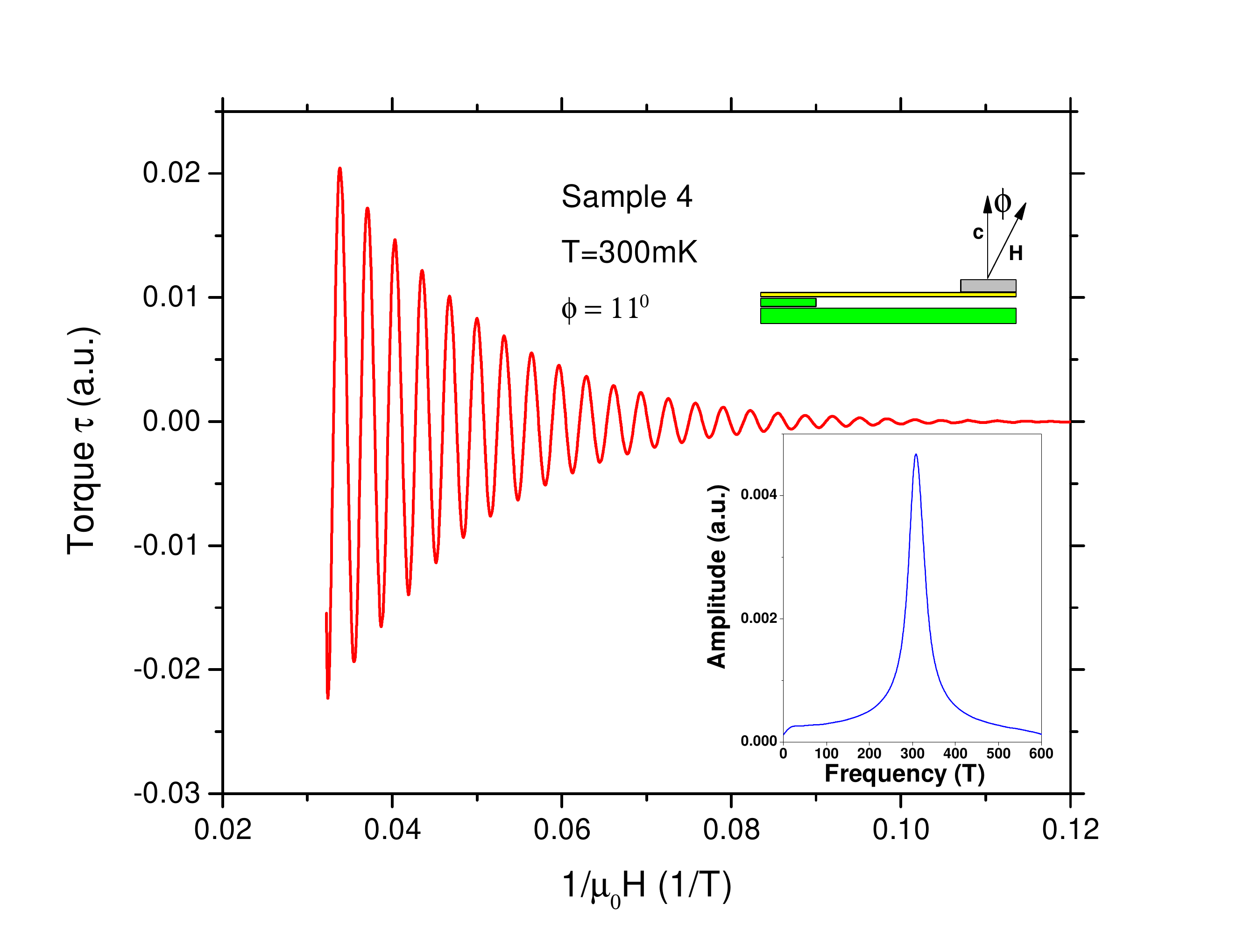}
\caption{\label{figTorque} (color online)
Oscillations in the torque data of Sample 4 with a polynomial background subtracted. The inset in the upper right hand corner is a schematic of the experimental setup. The lower inset is a Fast Fourier Transformation (FFT) of the oscillatory torque after subtracting the polynomial background. The single peak in the FFT spectrum reveals a single Fermi pocket.
}
\end{figure}

The dHvA effect was observed in all of our Cu$_x$Bi$_2$Se$_3$ crystals. A typical example of our torque data as a function of $1/\mu_{0}H$ with the polynomial background subtracted is shown in Fig. \ref{figTorque}. A single frequency of oscillations in magnetization reveal a single Fermi pocket. At 300 mK the oscillation frequency was measured as a function of angle up to 90$^{\circ}$. We further measured the temperature dependence of the oscillation amplitude at two different angles up to 25 K. A detailed analysis of the temperature and angular dependence of the quantum oscillations revealed the following features: 1) resolved quantum oscillations up to 90$^{\circ}$ at low carrier concentration shows a closed ellipsoidal Fermi surface, 2) the Fermi surface gets progressively elongated in the z-direction as carrier concentration increases and becomes open at high carrier density, 3) there is a strong effective mass anisotropy, and 4) the Fermi velocity is unchanged with increased carrier concentration supporting previous reports of a linear, Dirac-like dispersion in Cu$_x$Bi$_2$Se$_3$.

%%%%%%%%%%%%%%%%%%%%%%%%%%%%%%%%%%
%%%%%%%%%%%%%%%%%%%%%%%%%%%%%%%%%%
%%%%%%%%%%%%%%%%%%%%%%%%%%%%%%%%%% FIGURE 1

%%%%%%%%%%%%%%%%%%%%%%%%%%%%%%%%%%
%%%%%%%%%%%%%%%%%%%%%%%%%%%%%%%%%%

\begin{figure}[t]
\includegraphics[width= 3.5 in ]{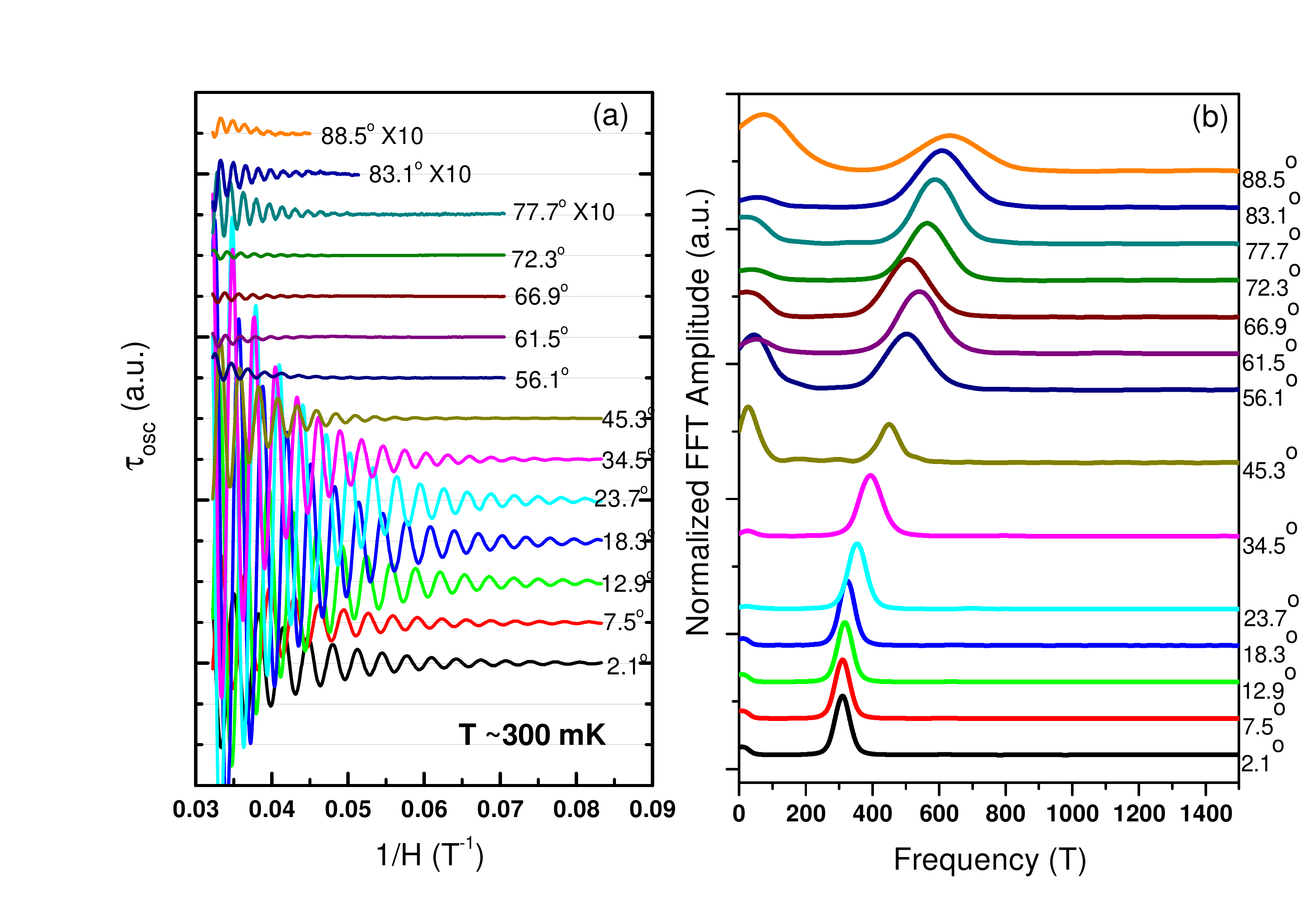}
\caption{\label{figangtorq} (color online)
(Panel a). Quantum oscillations in torque of Sample 4 at different angles after background subtraction. Oscillations are visible in the raw signal up to 90$^{\circ}$. At high tilt angle, the oscillation signal is multiplied by a factor of 10 for clarity. (Panel b). The FFT spectra of oscillations in panel a show a single Fermi pocket with clear angular dependence. The FFT amplitude is normalized by the height of the peak in the range of 200 T - 600 T. For the high tilt angles, the divergence of FFT amplitude in the DC end arises from an incomplete background subtraction. 
}
\end{figure}

\begin{figure}[t]
\includegraphics[width=3.5 in ]{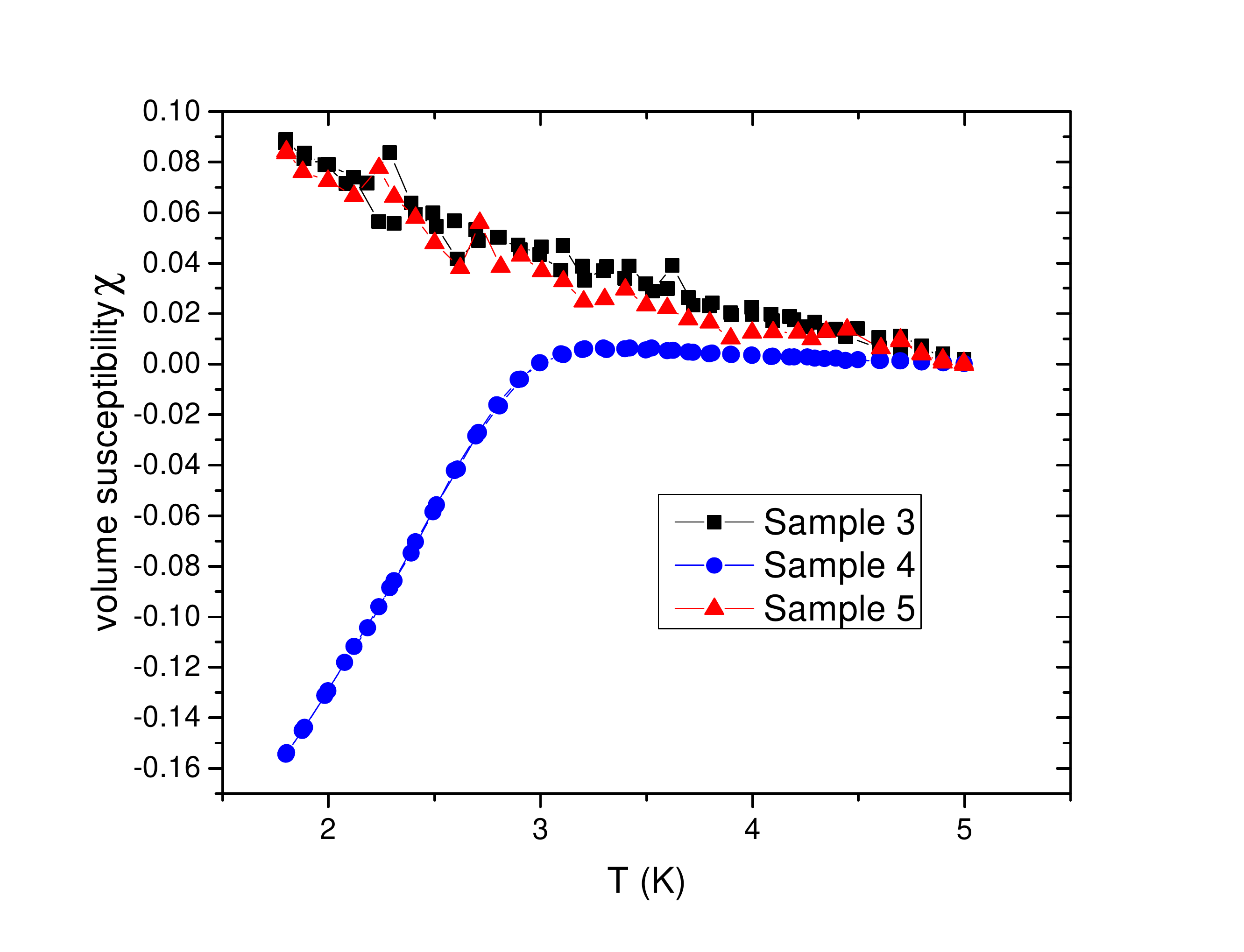}
\caption{\label{figsupercon} (color online)
Volume susceptibility measurements of 3 different samples. Sample 4, the sample with lowest carrier concentration, shows a superconducting transition at 3 K and a 16\% superconducting volume. Samples 3 and 5 do not show any superconducting property most likely due to sample quality degradation over time.
}
\end{figure}

Figure \ref{figangtorq} shows the torque signal from Sample 4. In Panel a, oscillations are clearly seen up to 90$^{\circ}$ in raw data indicating a closed Fermi surface. Panel b shows the FFT of the raw signal from panel a. Clear angular dependence can be tracked up to 90$^{\circ}$, where $H$ is parallel to the plane. Previous studies~\cite{Lawson}of the dHvA effect measured quantum oscillations up to 35$^{\circ}$. The observation of quantum oscillations up to 90$^{\circ}$ is important confirmation of the previous result that the Fermi Surface is an ellipsoid~\cite{Lawson}. We note the sample with this 3D Fermi surface is superconducting, and the Meissner effect of Sample 4 is shown in Fig. \ref{figsupercon} and discussed in detail there.

%%%%%%%%%%%%%%%%%%%%%%%%%%%%%%%%%%
%%%%%%%%%%%%%%%%%%%%%%%%%%%%%%%%%%
%%%%%%%%%%%%%%%%%%%%%%%%%%%%%%%%%% FIGURE 2

%%%%%%%%%%%%%%%%%%%%%%%%%%%%%%%%%%
%%%%%%%%%%%%%%%%%%%%%%%%%%%%%%%%%%

\begin{figure}[t]
\includegraphics[width= 4 in ]{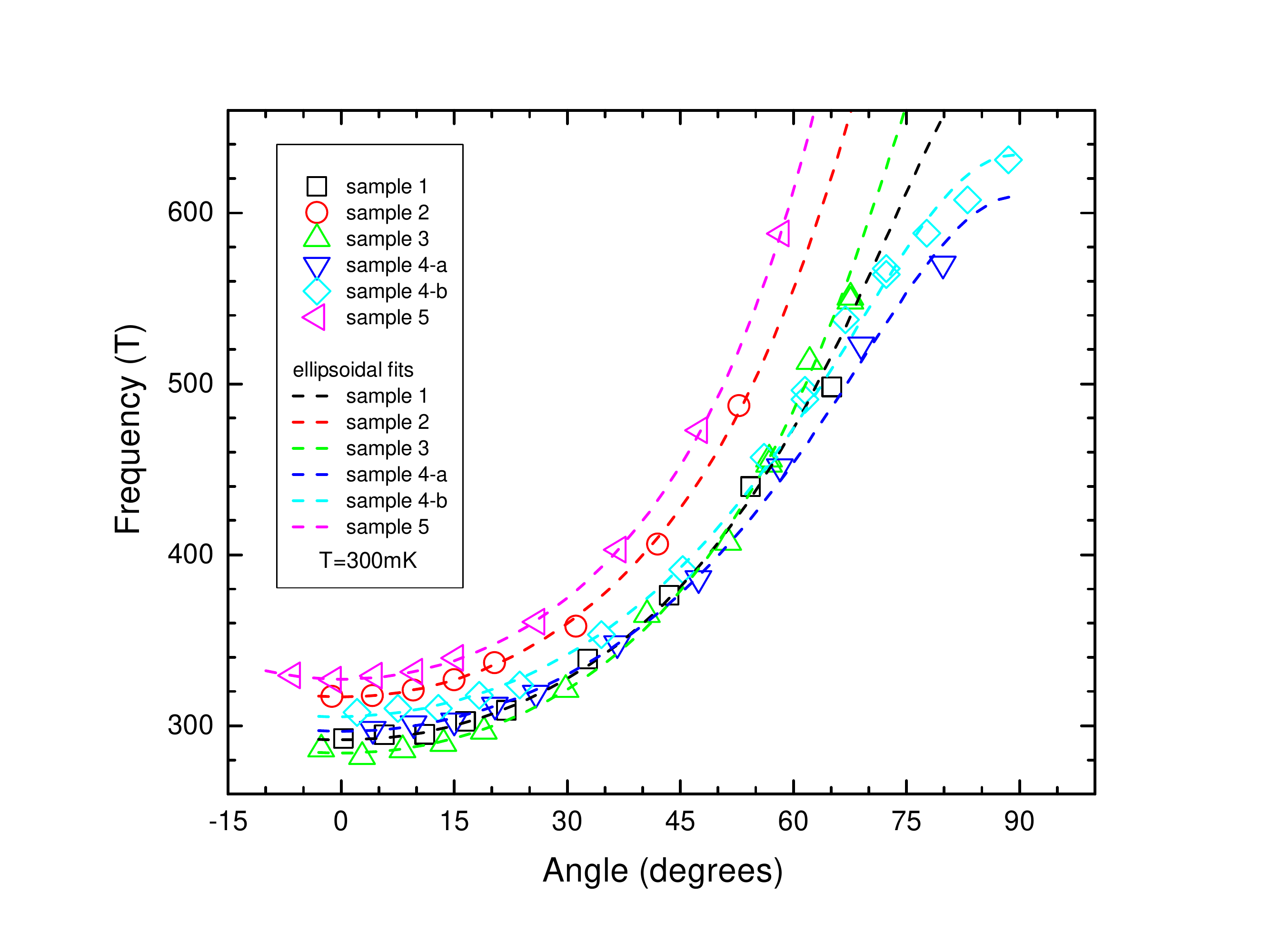}
\caption{\label{figAng} (color online)
Angular dependence of the oscillation frequency of the various samples. Dashed lines are ellipsoidal fits for the Fermi surfaces.
}
\end{figure}

Figure \ref{figAng} shows the angular dependence of the oscillation frequency for the various samples. The dashed lines are ellipsoidal fits given by $F(\phi)=F_{0}(\cos^{2}[\phi]+(\frac{k_{F}^{x}}{k_{F}^{z}})^{2}\sin^{2}[\phi])^{-\frac{1}{2}}$ where $F(\phi)$ is the frequency of the quantum oscillations at a particular $\phi$, and the fitted parameters are $F_{0}$ (the quantum oscillation frequency at $\phi$ = 0$^{\circ}$) and $\frac{k_{F}^{x}}{k_{F}^{z}}$ (a measure of the eccentricity of the Fermi surface). Most of the samples are fit well by a closed, ellipsoidal Fermi Surface; however, for the highest carrier concentration sample, a closed Fermi Surface fitting yields $k_{F}^{z}$ = 4.69 nm$^{-1}$, which is longer than the Brillouin Zone height~\cite{Kohler} of ~3.28 nm$^{-1}$. Thus, it is clear that the Fermi Surface becomes open at high carrier concentration - which was not seen in previous dHvA studies where there was only one sample of lower doping.~\cite{Lawson}

%%%%%%%%%%%%%%%%%%%%%%%%%%%%%%%%%%
%%%%%%%%%%%%%%%%%%%%%%%%%%%%%%%%%%
%%%%%%%%%%%%%%%%%%%%%%%%%%%%%%%%%% FIGURE 3

%%%%%%%%%%%%%%%%%%%%%%%%%%%%%%%%%%
%%%%%%%%%%%%%%%%%%%%%%%%%%%%%%%%%%

The angular dependence of the quantum oscillation frequency provides the size of the Fermi pocket. From the Onsager relation, the frequency of the quantum oscillation is proportional to the cross-section area  given by $A=\pi$$k_{F}^{x}$$k_{F}(\phi)$, with $k_F(\phi)$ and $k_x$ two semi-axes of the elliptical Fermi surface. Thus $F_{0}$ yields $k_{F}^{x}$ = $k_{F}^{y}$ and the eccentricity gives $k_{F}^{z}$. For Sample 4a,  $k_{F}^{x}$ = $k_{F}^{y}$ = 0.95nm$^{-1}$ and $\frac{k_{F}^{z}}{k_{F}^{x}}$ = 2.06. For a closed Fermi pocket, the bulk carrier concentration, $n$, is given by $n=\frac{1}{3\pi^{2}}k_{F}^{x}k_{F}^{y}k_{F}^{z}$. For the sample with the open Fermi surface, we calculated the bulk carrier concentration from finding the volume of the Fermi surface. This volume is arrived by integrating the ellipsoidal fit up to the Brillouin Zone boundary. We assume that the deviation from the ellipsoidal fit around the Brillouin Zone boundary due to bending is small. In this case the carrier concentration is given by $n=\frac{1}{2\pi^{2}}k_{F}^{x}k_{F}^{y}(k_{BZ}-\frac{k_{BZ}^{3}}{k_{F}^{z2}})$ where $k_{BZ}$ is the $\Gamma-Z$ distance. This yields a carrier concentration for Sample 5. The inferred carrier densities $n$ are listed with other electronic parameters in Table \ref{density}. 

%%%%%%%%%%%%%%%%%%%%%%%%%%%%%%%%%%%%%%%%%%%Table
\begin{table}[h]
\caption{\label{density} Summary of results in order of increasing carrier concentration. *The value of $k_{fz}/k_{fx}$ for sample 5 is ill-defined since $k_{fz}$ is taller than the Brillouin Zone. This is the value extracted from the ellipsoidal fit.}
\centering
\begin{ruledtabular}
\begin{tabular}{c c c c c}

                      					& $n (10^{19}cm^{-3})$		&$F_{0} (T)$		&$k_{fx} (nm^{-1})$		&$k_{fz}/k_{fx}$	\\
\hline
$4a$ 							&  5.93$\pm $0.24			&297$\pm $1 		&0.95$\pm $0.01		&2.06$\pm $0.05\\
$4b$							&  6.31$\pm $0.20			&306$\pm $1		&0.96$\pm $0.01		&2.09$\pm $0.02\\
$1$							&  6.78$\pm $0.73			&292$\pm $1		&0.94$\pm $0.01		&2.41$\pm $0.25\\
$3$							&  7.65$\pm $0.35			&284$\pm $1		&0.93$\pm $0.01		&2.83$\pm $0.09\\
$2$							&  10.05$\pm $1.18			&317$\pm $1 		&0.98$\pm $0.01		&3.16$\pm $0.36\\
$5$							&  13.91$\pm $2.71			&327$\pm $1		&1.00$\pm $0.01		&(4.69$\pm $0.79)*\\

\end{tabular}
\end{ruledtabular}
\label{parameter}
\end{table}
%%%%%%%%%%%%%%%%%%%%%%%%%%%%%%%%%%%%%%%%%%%table

The value of $k_{z}/k_{x}$ goes from 2.06 to 3.10 as the carrier concentration increases from 5.9x$10^{19}cm^{-3}$ to 10.1x$10^{19}cm^{-3}$ revealing that the Fermi Surface gets increasingly elongated in the z-direction as carriers are added. Then the Fermi Surface opens up and becomes quasi-cylindrical at high carrier concentration consistent with quantum oscillation measurements in magnetoresistence.~\cite{Lahoud}

\begin{figure}[t]
\includegraphics[width= 3.5 in ]{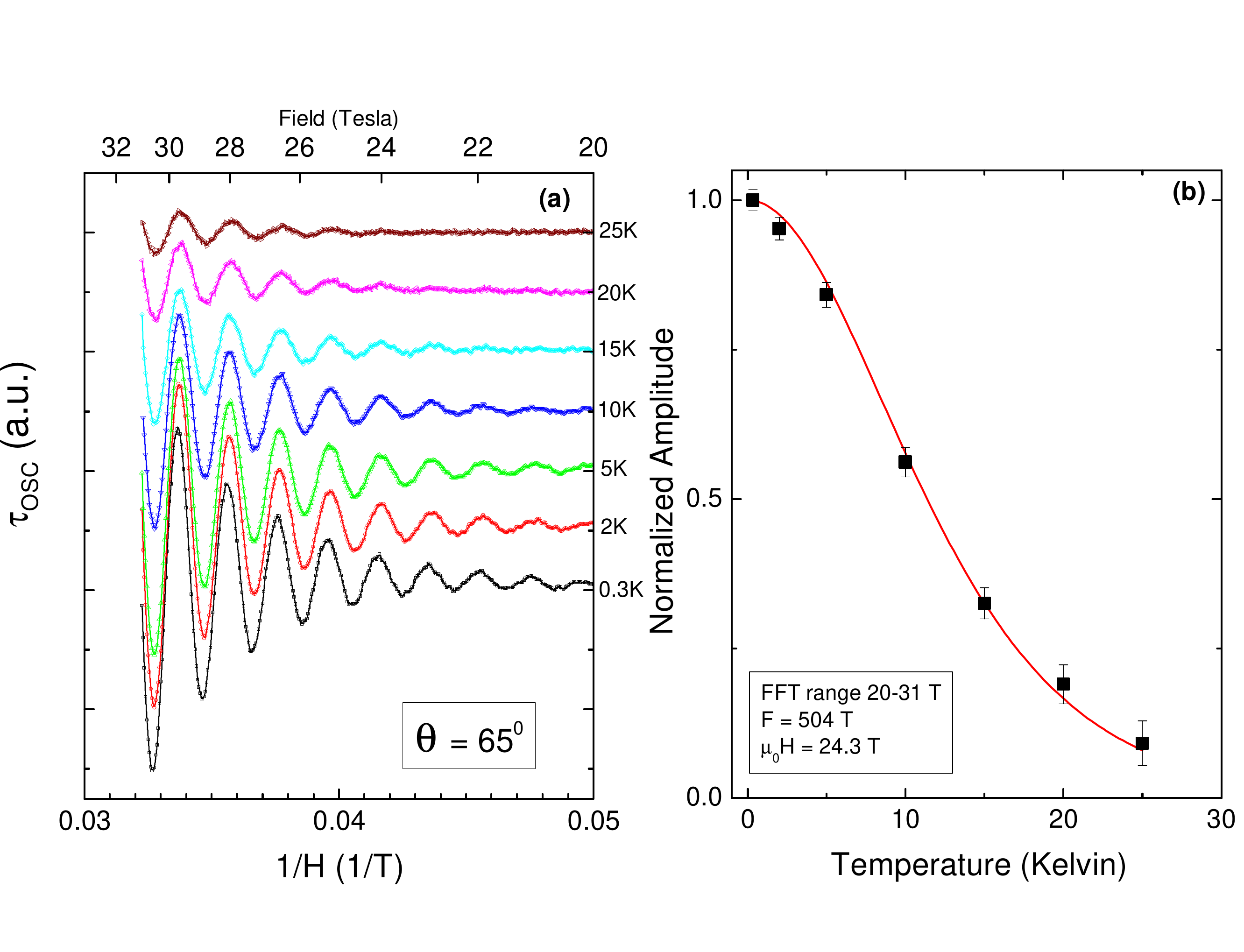}
\caption{\label{figTemp} (color online)
Panel a: Temperature dependence of the oscillation amplitude of Sample 4. Panel b: Temperature dependence of the normalized amplitude with fit to the thermal damping factor. Extracted from this fit is a high angle effective mass of 0.32m$_e$.
}
\end{figure}

The effective mass was extracted from the temperature dependence of the oscillation amplitudes. The amplitude of the dHvA oscillation is damped by the thermal damping factor~\cite{Shoenberg},
\begin{equation}
R_{T}=\frac{\alpha Tm^{*}}{B\sinh(\alpha Tm^{*}/B)}
\end{equation}
where the effective mass $m = m^*m_e$ and the Dingle temperature $T_D = \hbar/2\pi k_B \tau_S$. $\tau_S$ is the scattering rate, $m_e$ is the bare electron mass, $B = \mu_0H$ is the magnetic flux density, and  $\alpha = 2\pi^2 k_B m_e/e \hbar \sim $14.69 T/K. Panel a of Fig. \ref{figTemp} shows dHvA oscillations at temperatures ranging from 300 mK to 25 K. Panel b plots the normalized amplitudes of the peaks and the fitting is of the thermal damping factor from equation (2). For Sample 4, the effective mass increases from $0.16\pm0.01m_{e}$ to $0.32\pm0.01m_{e}$ as the angle increases from 15 to 65 degrees. Effective mass anisotropy was seen in very early studies of infrared reflection on Cu-doped Bi$_2$Se$_3$ measuring $m_{\parallel}/m_{\perp}$ to be ~4.35.~\cite{Tichy}

%%%%%%%%%%%%%%%%%%%%%%%%%%%%%%%%%%
%%%%%%%%%%%%%%%%%%%%%%%%%%%%%%%%%%
%%%%%%%%%%%%%%%%%%%%%%%%%%%%%%%%%% FIGURE 4

%%%%%%%%%%%%%%%%%%%%%%%%%%%%%%%%%%
%%%%%%%%%%%%%%%%%%%%%%%%%%%%%%%%%%

The Fermi velocity, $v_{F}$, was determined from the Fermi momentum, $k_{F}$, and the effective mass by $v_{f}= \hbar k_{F}/m$. The Fermi velocity is the slope of the energy momentum dispersion. For a quadratic dispersion, $v_{F}$ increases as $k_{F}$ gets larger. In contrast, for a linear dispersion, it should remain unchanged. In our results of a series of Cu doped Bi$_2$Se$_3$, the carrier density varies by an order of magnitude, yet the value of $v_F$ varies less than 16\%,  as shown in Table \ref{parameters}. Even though with copper doping the Fermi momentum increases from 0.69nm$^{-1}$ to 1.00nm$^{-1}$, the Fermi velocity remains relatively unchanged. The consistency of the Fermi velocity, is evidence for a linear, Dirac-like band in Cu$_x$Bi$_2$Se$_3$. This confirms this critical result observed in previous studies.~\cite{Wray, Lawson} Samples 3 and 4b are omitted from table \ref{parameters} since we only studied their temperature dependence at high angle and thus did not measure their in-plane effective mass.

%%%%%%%%%%%%%%%%%%%%%%%%%%%%%%%%%%%%%%%%%%%Table
\begin{table}[h]
\caption{\label{parameters} List of measured Fermi velocities of different samples in order of increasing Fermi momentum. The Fermi velocity remains unchanged indicating a linear dispersion. Last two samples are from Ref. ~\cite{Lawson}. Sample 4b and 3 are not listed here since the temperature dependence for those two samples were taken at high angle.}
\centering
\begin{ruledtabular}
\begin{tabular}{c c c c}

	                    			& $n (10^{19}cm^{-3})$	&$k_{fx} (nm^{-1})$			&$v_{f} (10^{5}m/s)$ 		\\
\hline
$1$						&  6.78				&  0.94					&6.4 					\\
$4a$ 						&  5.93				&  0.95					&6.9 					\\
$2$						&  10.05				&  0.98					&6.0 					\\
$5$						&  13.91				&  1.00					&6.8 					\\
Cu$_{0.25}$Bi$_2$Se$_3$		&  4.3				&  0.97					&5.8					\\
Bi$_2$Se$_3$				&  1.8				&  0.69					&5.7					\\

\end{tabular}
\end{ruledtabular}
\label{parameter}
\end{table}
%%%%%%%%%%%%%%%%%%%%%%%%%%%%%%%%%%%%%%%%%%%table

\begin{figure}[t]
\includegraphics[width=3.5 in ]{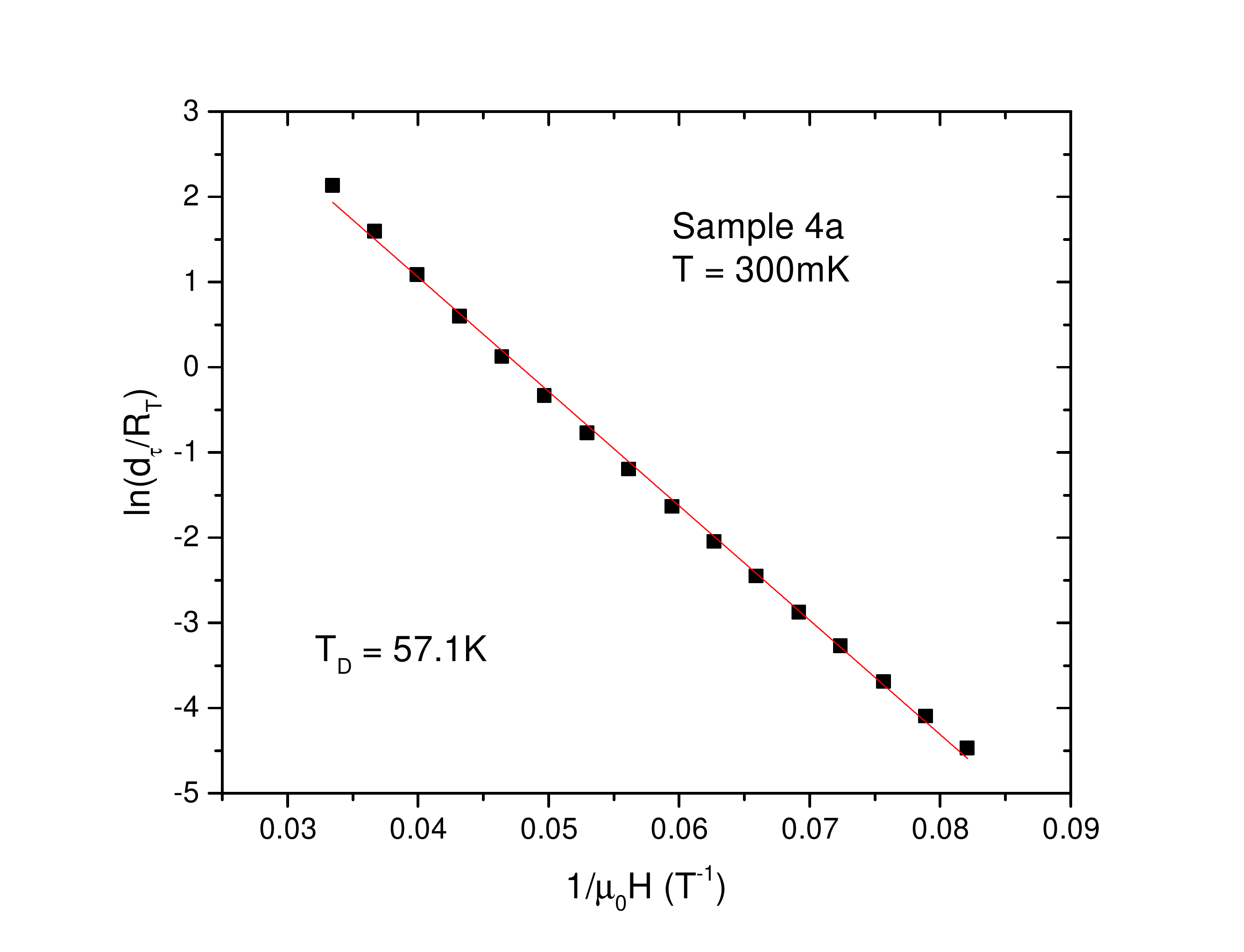}
\caption{\label{figDingle} (color online)
Dingle plot of Sample 4a. Fitting is of the Dingle damping factor and it yields a Dingle temperature of 57.1K.
}
\end{figure}

Further analysis of the quantum oscillation amplitude damping yields the mean free path and scattering rate of the samples. Figure \ref{figDingle} shows the Dingle plot of Sample 4. The fit of the Dingle damping factor, $R_{D}=exp[-\alpha T_{D}m/B]$, to this plot give a Dingle temperature~\cite{Shoenberg} of 57.1 K. $\alpha=2\pi^{2}k_{B}m_{e}/e\hbar\sim14.69\, T/K$. From the Dingle temperature, the scattering rate, $\tau$, can be extracted $T_{D}=\frac{\hbar}{2\pi k_{B}\tau}$. For this sample the scattering rate is 2.1x10$^{-14}$s. The mean free path is also determined from the scattering rate and the Fermi velocity by $l = v_{f} \tau$.

%%%%%%%%%%%%%%%%%%%%%%%%%%%%%%%%%%
%%%%%%%%%%%%%%%%%%%%%%%%%%%%%%%%%%
%%%%%%%%%%%%%%%%%%%%%%%%%%%%%%%%%% FIGURE 5

%%%%%%%%%%%%%%%%%%%%%%%%%%%%%%%%%%
%%%%%%%%%%%%%%%%%%%%%%%%%%%%%%%%%%

The results of the Dingle analysis including the scattering rate and mean free path of the various samples are given in Table \ref{Dingle}. Samples 3 and 4b are not shown in Table \ref{Dingle} since we only studied their temperature dependence at high angle and thus could not isolate the Dingle damping from the thermal damping at 0$^{\circ}$. Table \ref{Dingle} also shows the effective mass of the various samples for which temperature dependence was measured at low angle. Sample 4b and 3, for which temperature dependence was measured at high angle, have effective masses of 0.32m$_e$ and 0.29m$_e$ at 65$^{\circ}$ and 57$^{\circ}$ respectively.

With the exception of Sample 4, scattering rate and mean free path is relatively constant with added carriers varying by ~25\% in a random fashion. This is consistent with what was reported in the previous dHvA study.~\cite{Lawson} The average scattering rate, excluding the outlier, is 5.1x10$^{-14}$s and the average mean free path is 31nm. The variance in these parameters is based on sample quality. In the literature for clean samples, the Dingle temperature can be as low as 4 K~\cite{Analytis}, and for disordered samples, as high as 9.5 K.~\cite{Eto} This suggests a high level of disorder in our Cu$_{x}$Bi$_2$Se$_3$ samples - especially in Sample 4. 
%%%%%%%%%%%%%%%%%%%%%%%%%%%%%%%%%%%%%%%%%%%Table
\begin{table}[h]
\caption{\label{Dingle} Summary of results from effective mass and the Dingle analysis. Last two samples are from Ref. ~\cite{Lawson}. Sample 4b and 3 are not listed here since the temperature dependence for those two samples were taken at high angle.}
\centering
\begin{ruledtabular}
\begin{tabular}{c c c c c c}

                      					& $n (10^{19}cm^{-3})$		&$m^{*}/m_e$			&$T_{D} (K)$	&$\tau_{s} (10^{-14}sec)$		&$l (nm)$\\
\hline
$4a$ 							&  5.93					&0.16			&57.1 		&2.1						&15	\\
$1$							&  6.78					&0.17			&20.7		&5.9						&38	\\
$2$							&  10.05					&0.19			&25.8 		&4.7						&28	\\
$5$							&  13.91					&0.17			&27.7		&4.4						&30	\\
Cu$_{0.25}$Bi$_2$Se$_3$			&  4.3					&0.19			&23.5		&5.2						&30	\\	
Bi$_2$Se$_3$					&  1.8					&0.14			&23.9		&5.1						&29	\\

\end{tabular}
\end{ruledtabular}
\label{parameter}
\end{table}
%%%%%%%%%%%%%%%%%%%%%%%%%%%%%%%%%%%%%%%%%%%table

In addition to determining the electronic state, we further measured the the superconducting fraction of Cu doped Bi$_2$Se$_3$. The magnetic susceptibility was measured in an Quantum Design Magnetic Property Measurement System 2 weeks after the high field torque experiments. The sample with the lowest carrier concentration shows a superconducting transition with a 16\% superconducting volume as seen in Fig. \ref{figsupercon}. Two of the higher carrier concentration samples showed no superconducting transition suggesting either that the sample quality deteriorates with time and exposure or the superconducting phase is killed in the over-doped regime.

At high carrier concentration, the Fermi Surface becomes quasi-cylindrical and contains both the $\Gamma$ and Z points. This indicates that a topological superconducting state does not exist at high carrier concentration since the Fermi surface must enclose an odd number of time reversal invariant momenta in the Brillouin zone for a topological superconductor. However, since superconductivity coexists with the closed Fermi Surface (which can only contain the $\Gamma$ point) in the low carrier density sample, a topological superconducting state can still exist in the lower carrier density samples of Cu$_x$Bi$_2$Se$_3$. 

%%%%%%%%%%%%%%%%%%%%%%%%%%%%%%%%%%
%%%%%%%%%%%%%%%%%%%%%%%%%%%%%%%%%%
%%%%%%%%%%%%%%%%%%%%%%%%%%%%%%%%%% FIGURE 6

%%%%%%%%%%%%%%%%%%%%%%%%%%%%%%%%%%
%%%%%%%%%%%%%%%%%%%%%%%%%%%%%%%%%%

We note that Sample 4, though having the highest level of disorder, is the only sample to show a superconducting transition. The sample from the previous dHvA study had a Dingle temperature of 23.5 K much like the other samples and it also exhibited superconductivity.~\cite{Lawson} Therefore, it is not the case that superconductivity only occurs in the extreme case of highly disordered samples, rather we suggest that Sample 4 had undergone the least amount of degradation and just happened to be the sample with the highest disorder.

\section{Conclusion and Discussion}
Quantum oscillations in magnetization were resolved using highly sensitive torque magnetometry up to 31 T. A single Fermi pocket was observed to be increasingly elongated with added carriers. The effective mass has strong anisotropy, and the Fermi velocity remains unchanged with increasing Fermi momentum suggesting a linear, Dirac-like dispersion.

The nature of the transition of the Fermi surface topology is an interesting question. At higher concentration, the elongated 3D ellipsoidal Fermi surface touches the Fermi surface in the neighboring Brillouin Zone, mandating the transition from the 3D Fermi surface to a 2D quasi-cylindrical one.  Such an dramatic change of the Fermi surface topology suggests a Lifshitz transition as the Cu brings it extra carriers. Two experimental consequences are essential to confirm the dimensionality change and probe the nature of the transition. First, at higher $n$, the quasi-2D Fermi surface shall have two quantum oscillation frequencies, a large one from the belly, and a small one from the neck. The large frequency is what we observed in our dHvA measurement,~\cite{Lawson} and confirmed by the SdH measurements.~\cite{Lahoud} In contrast, the small neck frequency was not observed either in our dHvA measurements, nor in the SdH results.~\cite{Lahoud} This point is in particular puzzling, though Ref. ~\cite{Lahoud} reports that the signal arising from the small neck may be too small do to large effective mass effects. Therefore, the quantum oscillation measurements at dilution refrigerator temperature range is called to resolve the second oscillation frequency to confirm the 2D to 3D transition.

Another interesting experiment would be to the enhancement of thermopower near the 3D to 2D transition. A topology change in the electronic state usually leads to a large thermopower, a typical signature of Lifshitz transition.~\cite{Abrikosov, LiMg} Further thermoelectric measurements are essential to confirm this nature. If the dimensionality changes indeed occur and enhance the thermopower greatly, the Cu doping might lead to another interesting application of topological materials in thermoelectrics. 

{\bf Acknowledgement} We are grateful to discussion with Liang Fu, Kai Sun, and A. Kanigel,. The work is supported by the National Science Foundation under Award number ECCS-1307744 (low field torque magnetometry), the Department of Energy under Award number DE-SC0008110 (high field torque magnetometry), by the start up fund and the Mcubed project at the University of Michigan (low field magnetic susceptibility characterization), and by the National Science Foundation under Award number DMR-1255607 (sample growth).  The high-field experiments were performed at the National High Magnetic Field Laboratory, which is supported by NSF Cooperative Agreement No. DMR-084173, by the State of Florida, and by the DOE. We thank the assistance of Tim Murphy and Ju-Hyun Park of NHMFL. B. J. Lawson acknowledges the support by the National Science Foundation Graduate Research Fellowship under Grant No. F031543. T. Asaba thanks the support from the Nakajima Foundation.

\newpage

%%%%%%%%%%%%%%%%%%%%%%%%%%%%%%%%%%%%%%%%%%%
%%%%%%%%%%%%%%%%%%%%%%%%%%%%%%%%%%%%%%%%%%%
%%%%%%%%%%%%%%%%%%%%%%%%%%%%%%%%%%%%%%%%%%%

%

%
\end{document}